\documentclass[twocolumn,showpacs,preprintnumbers,amsmath,amssymb,superscriptaddress]{revtex4}
\usepackage{amssymb}
\usepackage{graphicx,color}
\usepackage{bm}

\newcommand \bef{\begin{figure}}
\newcommand \eef{\end{figure}}
\newcommand \befw{\begin{figure*}}
\newcommand \eefw{\end{figure*}}

\begin{document}

\title{Suppression of high $p_{T}$ hadrons in $Pb+Pb$ Collisions at LHC}

\date{\today}

\author{Xiao-Fang Chen}
\affiliation{Institute of Particle Physics, Huazhong Normal University, Wuhan 430079, China}

\author{Tetsufumi Hirano}
\affiliation{Department of Physics, The University of Tokyo, Tokyo 113-0033, Japan}
\affiliation{Nuclear Science Division, MS 70R0319, Lawrence Berkeley National Laboratory, Berkeley, CA 94720}

\author{Enke Wang}
\affiliation{Institute of Particle Physics, Huazhong Normal University, Wuhan 430079, China}

\author{Xin-Nian Wang}
\affiliation{Institute of Particle Physics, Huazhong Normal University, Wuhan 430079, China}
\affiliation{Nuclear Science Division, MS 70R0319, Lawrence Berkeley National Laboratory, Berkeley, CA 94720}

\author{Hanzhong Zhang}
\affiliation{Institute of Particle Physics, Huazhong Normal University, Wuhan 430079, China}

\begin{abstract}

Nuclear modification factor $R_{AA}(p_{T})$ for large transverse momentum pion spectra in
$Pb+Pb$ collisions at $\sqrt{s}=2.76$ TeV is predicted within the NLO perturbative QCD parton model.
Effect of jet quenching is incorporated through medium modified fragmentation functions within the
higher-twist approach. The jet transport parameter that controls medium modification is
proportional to the initial parton density and the coefficient is fixed by the RHIC data on suppression of large $p_{T}$
hadron spectra. Data on charged hadron multiplicity $dN_{ch}/d\eta=1584 \pm 80$ in central $Pb+Pb$ collisions
from the ALICE Experiment at the LHC are used to constrain the initial parton density both for
determining the jet transport parameter and the 3+1D ideal hydrodynamic evolution of the bulk matter
that is employed for the calculation of  $R_{PbPb}(p_{T})$ for neutral pions.

\end{abstract}


\maketitle

\section{Introduction}
One important evidence for the formation of strongly coupled quark-gluon
plasma (QGP) \cite{rhic1,rhic2,rhic3,rhic4} in high-energy heavy-ion collisions
at the Relativistic Heavy Ion Collider (RHIC) is the observation that the produced
dense matter is opaque to a propagating parton due to jet quenching \cite{Wang:1991xy}
which suppresses not only the single inclusive hadron spectra at large transverse
momentum \cite{star-suppression,phenix-suppression} but also back-to-back
high $p_{T}$ dihadron \cite{stardihadron} and $\gamma$-hadron
correlation \cite{Wang:1996yh,Adare:2009vd,Abelev:2009gu}.  These observed
jet quenching patterns in heavy-ion collisions at RHIC can be described by
perturbative QCD (pQCD) parton models well \cite{Wang:2004dn,Vitev:2002pf,Wang:2003mm,
Eskola:2004cr,Renk:2006nd,Zhang:2007ja,Qin:2007rn,Renk:2006qg,Zhang:2009rn,Qin:2009bk}
that incorporate parton energy loss \cite{Gyulassy:2003mc,Kovner:2003zj} as it propagates
through dense matter. Since the energy loss or medium modification of the effective jet fragmentation
functions depends on the space-time profile of  parton density in the medium, any systematic
and qualitative extraction of the properties of the medium through
phenomenological study of jet quenching has to take into account the dynamical
evolution of the bulk matter \cite{Bass:2008rv,Armesto:2009zi,Chen:2010te}.
The initial condition for the evolution of bulk matter is either
provided by model calculation or experimental data on the total charged hadron
multiplicity. Without any experimental data on hadron production in heavy-ion collisions
at the Large Hadron Collider (LHC), all predictions for jet quenching \cite{Abreu:2007kv}
have to rely on theoretical or phenomenological models on the initial condition for
bulk matter production and evolution.  However,
these theoretical and phenomenological predictions for the bulk hadron
production \cite{Abreu:2007kv} vary by almost a factor of two and, consequently, lead
to the same amount of uncertainties in the predictions for suppression of large transverse
momentum hadrons in heavy-ion collisions at the LHC.

Recently, ALICE Experiment at LHC published the first experimental data on the
charged hadron multiplicity density at mid-rapidity in central $Pb+Pb$ collisions at
$\sqrt{s} = 2.76$ TeV \cite{Aamodt:2010pb}. The measured $dN_{ch}/d\eta=1584 \pm 4 ({\rm stat.}) \pm 76 ({\rm sys.})$
for the top 5\% central $Pb+Pb$ collisions is larger than most of theoretical predictions,
especially those from the so-called color-glass-condensate models \cite{Aamodt:2010pb}.
Such an unexpected large hadron multiplicity will have important consequences on theoretical
predictions for jet quenching in $Pb+Pb$ collisions at the LHC energies.  The predictions for
single and dihadron suppression in the most central $Pb+Pb$ collisions at $\sqrt{s}=5.5$ TeV by
Wang, Wang and Zhang \cite{Abreu:2007kv} relied on the modified
HIJING calculation \cite{Li:2001xa} of the charged hadron multiplicity in heavy-ion collisions at the LHC.
The modification has now been incorporated into HIJING 2.0 version \cite{Deng:2010mv} of the original
HIJING1.0 model \cite{Wang:1991hta}. The new HIJINGH2.0 results \cite{Deng:2010mv, Deng:2010xg}
agree well with the first ALICE data \cite{Aamodt:2010pb} within experimental errors and theoretical uncertainty
which is controlled mainly by the uncertainty in nuclear shadowing of gluon distribution in nuclei.

With the first ALICE data \cite{Aamodt:2010pb} providing more stringent constraints on the theoretical
uncertainty in the bulk hadron production, we would like to revisit the predictions on suppression of
single inclusive hadron spectra at large $p_{T}$ in heavy-ion collisions at LHC. Moreover, we will
use the 3+1D ideal hydrodynamic model for more realistic description of the bulk matter evolution
whose initial conditions at LHC are also much better constrained by the first ALICE data.

The rest of the paper is organized as follows. In the next section, we will provide a brief overview
of the pQCD parton model for single inclusive hadron spectra and the high-twist (HT) approach to
the medium modified fragmentation functions. In Sec. III, we will give a brief description of the
3+1D ideal hydrodynamic model for the bulk evolution with initial condition provided by HIJING2.0
model and further constrained by the first ALICE data. In Sec. IV, we present predictions
for the nuclear modification factor $R_{AA}(p_{T})$ for neutral pions in $Pb+Pb$ collisions at LHC
energy $\sqrt{s}=2.76$ TeV/$n$ and discuss about first LHC data for the charged hadrons from the
ALICE Experiment. We will give a summary and discussion in Sec. V.

\section{pQCD parton model and medium-modified fragmentation functions}

We will employ the next leading order (NLO) pQCD parton model for the initial
jet production spectra which has been shown to work well for large
$p_T$ hadron production in high energy nucleon-nucleon
collisions~\cite{Owens}. We use the same factorized form for the inclusive
particle production cross section in $A+B$ heavy-ion  collisions, which
can be expressed as a  convolution of parton distribution functions inside
the nuclei (nucleons),  elementary parton-parton scattering cross sections
and effective parton fragmentation functions,
\begin{eqnarray}
\frac{d\sigma^h_{AB}}{dyd^2p_T}&=&\sum_{abcd}\int d^{2}b d^2r
dx_adx_b t_A({\bf r})t_B(|{\bf r}-{\bf b}|) \nonumber \\
&&\hspace{-0.1in}\times\
f_{a/A}(x_a,\mu^2) f_{b/B}(x_b,\mu^2) \frac{d\sigma}{d\hat{t}}(ab\rightarrow cd)\nonumber \\
&&\hspace{-0.1in}\times \frac{\widetilde D_{h/c}(z_{c},\mu^2,E, b, r)}{\pi z_{c}} +\mathcal {O}(\alpha_s^3),
\label{eq:AA}
\end{eqnarray}
where $d\sigma(ab\rightarrow cd)/d\hat{t}$ are elementary parton
scattering cross sections at leading order (LO) $\alpha_s^2$. The average over the
azimuthal angle of the initial jet is implicitly implied in the above equation since we focus on
azimuthal integrated single inclusive hadron cross section. The NLO contributions include
$2\rightarrow 3$ tree level contributions and 1-loop virtual corrections
to $2\rightarrow2$ tree processes \cite{Owens}. The nuclear thickness function
is normalized to $\int d^{2}r t_{A}({\bf r})=A$. The nuclear parton
distributions per nucleon $f_{a/A}(x_a,\mu^2,{\bf r})$ are assumed
to be factorized into the parton distributions in a free nucleon $f_{a/N}(x,\mu^2)$
and the nuclear shadowing factor $S_{a/A}(x,\mu^2,{\bf r})$,
\begin{eqnarray}
   f_{a/A}\left(x,\mu^2,\mathbf{r}\right)&=&S_{a/A}\left(x,\mu^2,\mathbf{r}\right)
  \left[\frac{Z}{A}f_{a/p}\left(x,\mu^2\right)\right.
 \nonumber \\
&~& \left.+\left(1-\frac{Z}{A}\right)f_{a/n}\left(x,\mu^2\right)
\right],\label{eqn:shad}
\end{eqnarray}
where $Z$ is the  charge and $A$ the mass number of the nucleus. The CTEQ6M parameterization \cite{distribution}
for parton distribution functions will be used for nucleon parton distributions $f_{a/N}(x,\mu^2)$.
The parton shadowing factor $S_{a/A}(x,\mu^2,\mathbf{r})$ describes the nuclear modification
 of parton distributions per nucleon inside a nucleus and can be given by
parameterizations \cite{EPS08}. With an impact-parameter-dependent parton shadowing the effect of shadowing
is the strongest for hard scatterings at the center of the transverse plane. But partons from these hard scatterings
at the center are mostly quenched due to parton energy loss and do not contribute to the final hadron spectra.
Therefore, the effect of parton shadowing is negligible for the final hadron spectra in $A+A$ collisions that are dominated
by hard scattering close to the surface of the overlapping nuclei \cite{wwzshadow}.
We will set  $S_{a/A}(x,\mu^2,\mathbf{r})$=1.0 in the calculation.
Effect of jet quenching in dense medium in heavy-ion collisions will be described by effective medium modified
parton fragmentation functions $\widetilde{D}_{c}^{h}(z_{h},Q^2,E,b,r)$ within
the HT approach.

Within the generalized factorization of twist-four processes, one can calculate the parton energy loss and
medium modified fragmentation functions of a propagating parton in the medium after it is produced via a hard
process \cite{guoxiaofeng,benwei-nuleon}. Within such a high-twist approach,
the medium modification to the parton fragmentation functions is caused by
multiple scattering inside the medium and  induced gluon bremsstrahlung.
The medium modified quark fragmentation functions,
\begin{eqnarray}
\tilde{D}_{q}^{h}(z_h,Q^2) &=&
D_{q}^{h}(z_h,Q^2)+\frac{\alpha_s(Q^2)}{2\pi}
\int_0^{Q^2}\frac{d\ell_T^2}{\ell_T^2} \nonumber\\
&&\hspace{-0.7in}\times \int_{z_h}^{1}\frac{dz}{z} \left[ \Delta\gamma_{q\rightarrow qg}(z,\ell_T^2)D_{q}^h(\frac{z_h}{z})\right.
\nonumber\\
&&\hspace{-0.2 in}+ \left. \Delta\gamma_{q\rightarrow
gq}(z,\ell_T^2)D_{g}^h(\frac{z_h}{z}) \right] ,
\label{eq:mo-fragment}
\end{eqnarray}
take a form very similar to the vacuum bremsstrahlung corrections that lead to the
evolution equations in pQCD for fragmentation
functions, except that the medium modified splitting functions,
$\Delta\gamma_{q\rightarrow qg}(z,\ell_T^2)$ and
$\Delta\gamma_{q\rightarrow gq}(z,\ell_T^2)=\Delta\gamma_{q \rightarrow qg}(1-z,\ell_T^2)$
depend on the properties of the medium via the jet transport
parameter $\hat q$ \cite{Chen:2010te,CasalderreySolana:2007sw},
\begin{equation}
\hat q=\rho \int dq_T^2 \frac{d\sigma}{dq_T^2} q_T^2.
\label{eq:qhat0}
\end{equation}
or the average squared transverse momentum broadening per unit
length, which is also related to the gluon distribution density of the
medium \cite{CasalderreySolana:2007sw,Baier:1996sk}.
The corresponding quark energy loss can be expressed as \cite{Chen:2010te,CasalderreySolana:2007sw},
\begin{eqnarray}
\frac{\Delta E}{E} &=& \frac{2N_{c}\alpha_s}{\pi} \int dy^-dz
{d\ell_T^2}
\frac{1+z^2}{\ell_T^4} \nonumber \\
&& \hspace{-0.5in}\times \left(1-\frac{1-z}{2}\right)\hat q(E,y)
\sin^2\left[\frac{y^-\ell_T^2}{4Ez(1-z)}\right], \label{eq:de-twist}
\end{eqnarray}
in terms of the jet transport parameter.  Note that we include an extra factor of $1-(1-z)/2$ as compared to that
used in Refs. \cite{CasalderreySolana:2007sw,Deng:2009qb} due to corrections beyond the helicity
amplitude approximation \cite{benwei-nuleon}. We refer readers to Ref.~\cite{Chen:2010te}
for details of the modified fragmentations. The fragmentation functions $D^0_{h/c}(z_c,\mu^2)$
in the vacuum  are given by the  updated AKK parametrization~\cite{akk08}.

According to the definition of jet transport parameter, we can assume it to be proportional
to local parton density in a QGP and hadron density in a hadronic gas. Therefore, in
a dynamical evolving medium, one can express it in general as
\begin{equation}
\label{q-hat-qgph}
\hat{q} (\tau,r)= \left[\hat{q}_0\frac{\rho_{QGP}(\tau,r)}{\rho_{QGP}(\tau_{0},0)}
  (1-f) + \hat q_{h}(\tau,r) f \right]\cdot \frac{p^\mu u_\mu}{p_0}\,,
\end{equation}
where $\rho_{QGP}$ is the parton (quarks and gluon) density in an ideal gas at a given temperature,
$f(\tau,r)$ is the fraction of the hadronic phase at any given
space and time, $\hat q_{0}$ denotes the jet transport
parameter at the center of the bulk medium in the QGP phase at the
initial time $\tau_{0}$, $p^\mu$ is the four momentum of the jet and $u^\mu$ is the
four flow velocity in the collision frame. We assume the hadronic phase of the medium
is described as a hadron resonance gas, in which the jet transport parameter is approximated as,
\begin{equation}
\hat q_{h}=\frac{\hat q_{N}}{\rho_{N}}\left[ \frac{2}{3}\sum_{M}\rho_{M}(T)+\sum_{B}\rho_{B}(T)\right],
\label{eq:qhath}
\end{equation}
where $\rho_{M}$ and $\rho_{B}$ are the meson and baryon density in the hadronic resonance gas at
a given temperature, respectively, $\rho_{N}=n_{0}\approx 0.17$ fm$^{-3}$ is the nucleon density in the
center of a large nucleus and the
factor $2/3$ accounts for the ratio of constituent quark numbers in mesons and baryons.
The jet transport parameter at the center of a large nucleus $\hat q_{N}$ has been studied in
deeply inelastic scattering (DIS) \cite{Wang:2002ri,Majumder:2004pt,Majumder:2009zu} .
We use a recently extracted value \cite{Deng:2009qb}  $\hat q_{N}\approx 0.02$ GeV$^{2}$/fm
from the HERMES \cite{Airapetian:2007vu} experimental data.
The hadron density at a given temperature $T$ and zero chemical potential is
\begin{eqnarray}
\rho_{h}(T)=\frac{T^{3}}{2\pi^{2}}
\left(\frac{m_{h}}{T}\right)^{2}
\sum_{n=1}^{\infty}\frac{\eta_{h}^{n+1}}{n}K_{2}\left(n\frac{m_{h}}{T}\right),
\end{eqnarray}
where $\eta_{h}=\pm$ for meson (M)/baryon (B). In the paper, we will include all hadron resonances with
mass below 1 GeV.

\section{3+1D ideal hydrodynamic evolution of bulk matter}

In the model for medium modified fragmentation functions as described in the last section,
one needs information on the space-time evolution of the local temperature and flow velocity
in the bulk medium along the jet propagation path. We will use a full three-dimensional 3+1D ideal
hydrodynamics \cite{Hirano2001,HT2002} in our calculation to describe the space-time evolution
of the bulk matter in heavy-ion collisions.

We solve equations of energy-momentum conservation
in full 3+1D space $(\tau,x,y,\eta_s)$ under the assumption that
local thermal equilibrium is reached at an initial
time $\tau_{0}$ =0.6 fm/$c$ and
maintained thereafter until freeze-out. Here $\tau$, $\eta_s$,
$x$ and $y$ are proper time, space-time rapidity, and
two transverse coordinates perpendicular to the beam axis, respectively.
Ideal hydrodynamics is characterized by the energy-momentum tensor,
\begin{equation}
T^{\mu\nu}=(\epsilon+P)u^{\mu}u^{\nu}-Pg^{\mu\nu},
\end{equation}
where $\epsilon$, $P$ and $u^{\mu}$ are energy density, pressure and
local four velocity, respectively. We neglect the finite net-baryon
density which is supposed to be small both at RHIC and LHC energies.
For the quark-gluon plasma (QGP) phase at high temperature ($T>T_{c}=170$ MeV),
we use the equation of state (EOS) of a relativistic massless parton gas
 ($u$, $d$, $s$ quarks and gluons) with a bag pressure $B$:
\begin{eqnarray}
p=\frac{1}{3}(\epsilon-4B).
\end{eqnarray}
The bag constant is tuned to be $B^{\frac{1}{4}}=247$\,MeV to
match the pressure of the QGP phase to that of a hadron resonance gas
at critical temperature $T_{c}=170$\,MeV.
Below the critical temperature $T<T_{c}$, a hadron resonance gas
model including all hadrons up to the mass of the $\Delta(1232)$ is employed for the EOS .
The hadron resonance gas EOS employed in this study
implements chemical freeze-out
at $T_{\mathrm{ch}}=170$ MeV \cite{HT2002}
so that evolution of
chemically frozen, but thermally equilibrated hadronic
matter is described. In the calculation of parton energy loss and medium
modified fragmentation functions, jets propagate along a straight path
in the evolving medium until the decoupling of the medium at $T^{\mathrm{dec}} = 100$ MeV.

\begin{figure}[htb]
\includegraphics[width=3.3in]{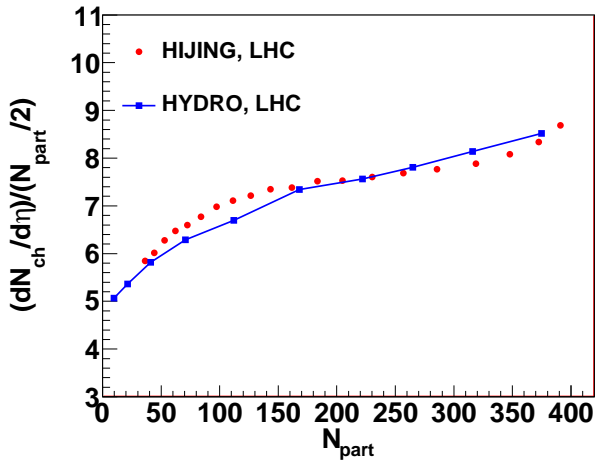}
\caption{(Color online)
Charged particle multiplicity density at midrapidity per
participant pair as a function of the number of participants
in $Pb+Pb$ collisions at $\sqrt{s_{NN}}$ = 2.76 GeV
from HIJING 2.0 (solid circles) \cite{Deng:2010mv} and ideal hydrodynamics (solid square). }
\label{fig:nchnpart}
\end{figure}

For the initial condition of the longitudinal flow velocity,
Bjorken's scaling solution \cite{Bjorken} is employed.
The initial entropy distribution in the transverse plane is proportional
to a linear combination of the number density of participants, $\rho_{\mathrm{part}}$,
and that of binary collisions, $\rho_{\mathrm{coll}}$~\cite{HHKLN2006}.
The proportionality constant and the fraction of soft and hard components
are so chosen that the centrality dependence of charged particle multiplicity
agrees with the experimental data at RHIC \cite{HHKLN2006} and HIJING2.0 results
at the LHC energy \cite{Deng:2010mv} which agree with the recent ALICE data for the most
central $Pb+Pb$ collisions at $\sqrt{s}=2.76$ TeV \cite{Aamodt:2010pb}. Shown
in Fig.~\ref{fig:nchnpart} is the hydrodynamic calculation (solid square) of the charged hadron multiplicity density
at mid-rapidity as compared with the HIJING2.0 result (solid circle).

%
%

We will use these hydrodynamic solutions to provide values of temperature
and flow velocity along the trajectory of parton propagation for the evaluation of local jet transport
parameter in Eq.~(\ref{q-hat-qgph}) for
the calculation of medium modified fragmentation functions.
The hydrodynamic results obtained in this way such as temperature
and energy density at each space-time position are publicly available \cite{hydro-site}.
Note that these hydrodynamic results
provide the evolution of the space-time profiles of  jet transport parameter and gluon density relative
to their values at the center of overlapped region of dense matter ($r=0$) in the most central collisions ($b=0$)
at an initial time $\tau_{0}$. The normalization of the initial values at $\tau_{0}$ and $r=0, b=0$ is included in
the value of $\hat q_{0}$. It should be proportional to the initial parton density, which in turn is determined by fitting
the final charged hadron multiplicity density in mid-rapidity (see Fig.~\ref{fig:nchnpart}) of the hydrodynamics results.
Therefore, approximately,
\begin{equation}
\hat q_{0}\propto \frac{1}{\pi \tau_{0}R_{A}^{2}}\frac{dN_{ch}}{d\eta}.
\label{qhat0}
\end{equation}
We determine the coefficient of the above relation by fitting the experimental data at the RHIC energy for
the most central $Au+Au$ collisions. Then the energy dependence and the impact-parameter
dependence will be completely determined by the measurement or calculated values of charged hadron multiplicity
density $dN_{ch}/d\eta$.

\subsection{ Numerical results}

We use a NLO Monte Carlo based program~\cite{Owens} to calculate the
single hadron spectra in our study. In this NLO program,  the
factorization scale and the renormalization scale are chosen to be the
same (denoted as $\mu$) and are all proportional to the transverse momentum of the final hadron $p_T$.
We choose $\mu=1.2p_T$ with which the calculated single inclusive pion spectra for $p+p$ collisions
in NLO pQCD agree well with RHIC data \cite{Zhang:2007ja}. We will use the same scale in $A+A$
collisions at LHC energies.

With space-time profile of the gluon density provided by the hydrodynamic evolution equations
we calculate medium modified fragmentation functions  which are used then to calculate
the suppression factor (or nuclear modification factor) for large $p_{T}$ hadron spectra in
heavy-ion collisions \cite{Wang:1998ww},
\begin{eqnarray}
R_{AB}=\frac{d\sigma_{AB}^h/dyd^2p_T}{N_{bin}^{AB}(b)
d\sigma_{pp}^h/dyd^2p_T}, \label{eq:rab}
\end{eqnarray}
where $N_{bin}^{AB}(b)=\int d^{2}r t_{A}(r)t_{B}(|\vec b-\vec r|)$. The fixed value of impact-parameters
in the calculation of the spectra and the modification factor are determined through the Glauber
geometric fractional cross sections for given centrality of the heavy-ion collisions.

\begin{figure}[t]
\begin{center}
\includegraphics[width=90mm]{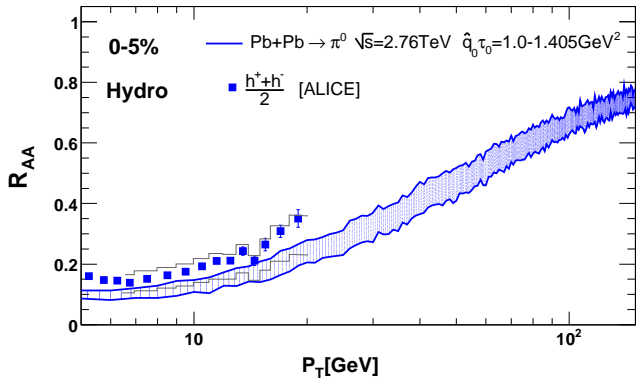}
\caption{ (Color Online) Nuclear modification factor at mid-rapidity for neutral pion spectra in
the most  $0-5\%$ central $Pb+Pb$ collisions at $\sqrt{s}=2.76$ TeV, using the HT modified fragmentation functions with
$\hat q_{0}\tau_{0}=1.0-1.4$ GeV$^2$. The data points (filled square) are for charged hadrons in the same central $Pb+Pb$
collisions from ALICE Experiment \cite{Aamodt:2010jd} with the histogram representing systematic errors due to uncertainty
in the $p+p$ spectra at $\sqrt{s}=2.76$ TeV from interpolation.}
\label{fig-ht}
\end{center}
\end{figure}

In the HT approach, we have determined $\hat q_{0}\tau_{0}=0.54-0.63$ GeV$^2$ from the experimental data
on pion spectra in the most 0-10\% central $Au+Au$ collisions at $\sqrt{s}=0.2$ TeV \cite{Chen:2010te}.
We assume that the jet transport parameter is proportional to the initial parton density or the transverse density of
charged hadron multiplicity in mid-rapidity [Eq.~\ref{qhat0}]. With the new ALICE data on
charged particle pseudo-rapidity density at mid-rapidity $dN_{ch}/d\eta=1584\pm 4 (stat.) \pm 76 (sys.)$ \cite{Aamodt:2010pb}
 in the most central $5\%$ $Pb+Pb$ collisions at $\sqrt{s}=2.76$ TeV versus $dN_{ch}/d\eta=687\pm 37$ for
 0-5\% $Au+Au$ collisions at $\sqrt{s}=0.2$ TeV \cite{phenix-nch}, we obtain the extrapolated
 value $\hat q_{0}\tau_{0}=1.0-1.4$ GeV$^{2}$ for $Pb+Pb$ collisions at $\sqrt{s}=2.76$ TeV.
 Shown in Fig.~\ref{fig-ht} is the predicted nuclear modification factor for pion spectra in the 0-5\%
  central $Pb+Pb$ collisions at $\sqrt{s}=2.76$ TeV. The suppression factor increases
with $p_{T}$ partially because of the energy dependence of parton energy loss and partially because of the less
steep initial jet production spectra \cite{Wang:2004yv}. This trend is similar to almost all LHC predictions by many
other parton energy loss model calculations \cite{Abreu:2007kv}.

 We also show in Fig.~\ref{fig-ht} the recently published ALICE data \cite{Aamodt:2010jd} (filled square) on the suppression factor
 for charged hadrons in the most 0-5\% central $Pb+Pb$ collisions at the LHC energy $\sqrt{s}=2.76$ TeV.  Since there are no
 experimental data on charged hadron spectra in $p+p$ collisions at the LHC energy $\sqrt{s}=2.76$ TeV,  ALICE data on the
 suppression factor were obtained with $p+p$ spectra interpolated from experimental data at Tevatron energies. The histograms
 in Fig.~\ref{fig-ht} represent the errors on the suppression factor from the uncertainty in the interpolation. Charged hadrons
 contain significant fraction of protons and anti-protons which could have nonnegligible contributions from parton
 recombination \cite{Hwa:2002tu,Greco:2003xt,Fries:2003vb}. Because of the abundance of jet production
 at the LHC energy, recombination among these hard partons becomes possible and therefore contribute to hadron, especially
 baryon spectra at high $p_{T}$. One therefore should take into account the contribution from hard parton recombination
 in the calculation of the final charged hadron spectra, which could push the suppression factor for charged hadrons higher
 than that for pions at large $p_{T}$.

\begin{figure*}[t]
\begin{center}
\includegraphics[width=6in]{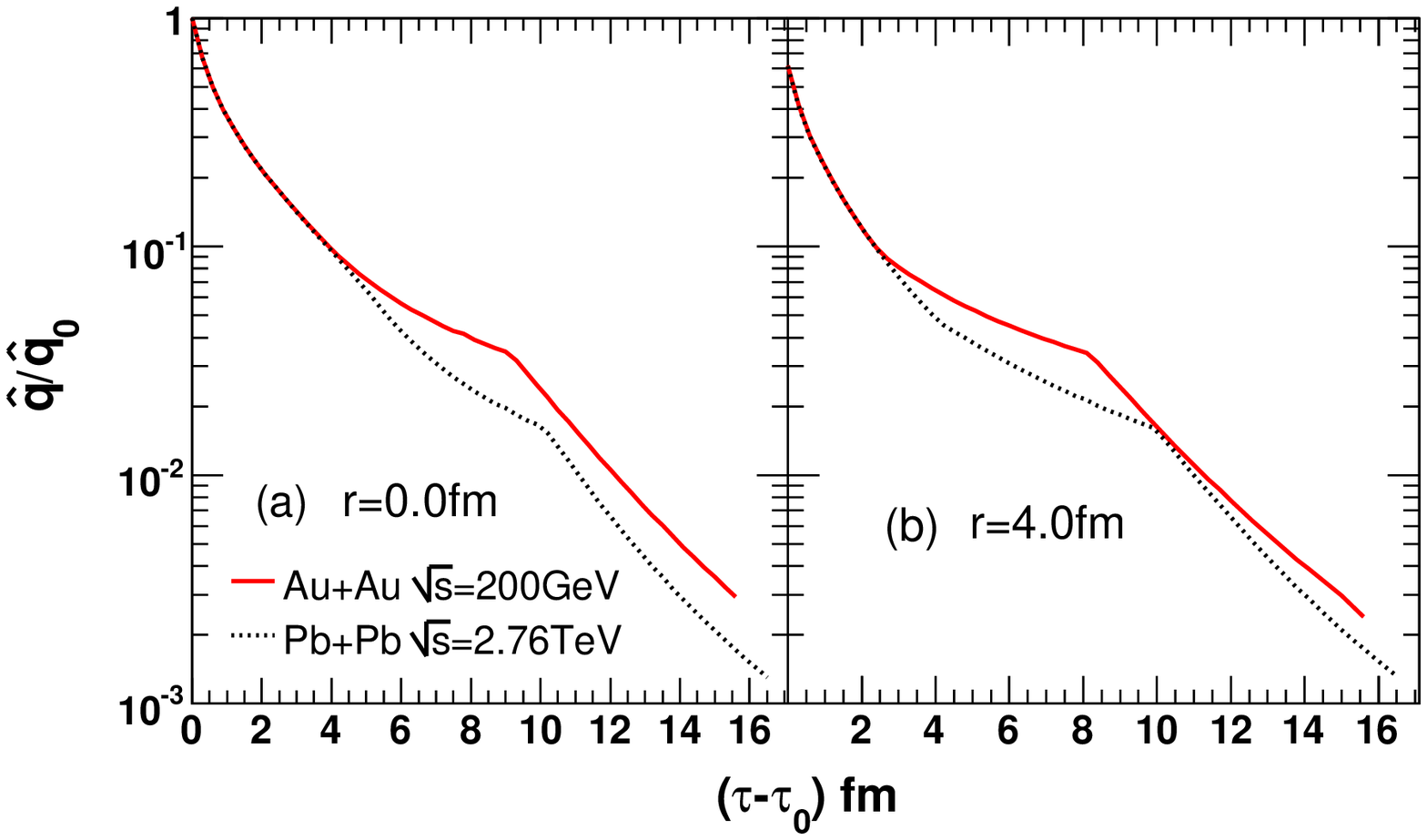}
\caption{(Color Online) The scaled jet transport parameter $\hat q(r,\tau)/\hat q(0,\tau_{0})$ as a function of $\tau-\tau_{0}$
at $r=0$ fm and $r=4$ fm in central $Au+Au$ collisions at RHIC ($\sqrt{s}=0.2$ TeV) and $Pb+Pb$ collisions at LHC ($\sqrt{s}=2.76$ TeV)
from 3+1D hydrodynamic evolution.}
\label{fig-rho}
\end{center}
\end{figure*}

Because of the increased initial parton density which we assume to be proportional to the final hadron multiplicity
density, the initial jet transport parameter $\hat q_{0}$  in $Pb+Pb$ collisions at the LHC
energy $\sqrt{s}=2.76$ TeV are more than twice larger than that in $Au+Au$ collisions at the RHIC energy $\sqrt{s}=0.2$ TeV.
The hadron suppression factors in $Pb+Pb$ collisions at LHC at moderate transverse momentum $p_{T}<20$ GeV/$c$
are therefore about twice smaller than that at RHIC \cite{Chen:2010te}. Shown in Fig.~ \ref{fig-rho}
are the scaled jet transport parameter $\hat q(r,\tau)/\hat q(0,\tau_{0})$ as a function of $\tau-\tau_{0}$ which is
related to the parton and hadron density [Eq.~(\ref{q-hat-qgph})] as given by the hydrodynamic evolution. The kink
in the time dependence is caused by the first-order phase transition assumed in the hydrodynamics evolution as the EOS
of the dense matter. For most part of the evolution history, the scaled jet transport parameters are very similar
at RHIC and LHC energies. Therefore, the increased hadron suppression at LHC is caused mainly by the overall
increase of the initial parton density. The increased initial parton density, however, will also increase the life-time
of the dense matter through-out the phase transition and hadronic phase. This will also contribute to the increased
suppression of hadron spectra at the LHC as compared to at RHIC.

\section{Conclusions}
\label{sec:c}

We used the new ALICE data on charged hadron multiplicity density at mid-rapidity in central $Pb+Pb$ collisions at
the LHC energy $\sqrt{s}=2.76$ TeV \cite{Aamodt:2010pb} to estimate the initial jet quenching parameters in $Pb+Pb$
collisions at the LHC and the initial condition for the hydrodynamic evolution of the bulk matter. With the initial
values of the jet transport parameter and the initial condition for hydrodynamic evolution of the bulk matter, we predict the suppression
factor for the hadron spectra in $Pb+Pb$ collisions at $\sqrt{s}=2.76$ TeV within the HT model for
medium modified fragmentation functions. Because of the increased initial parton density of about a factor of two,
and the longer life-time of the dense matter or the duration of jet quenching, the hadron spectra are found to be suppressed
more at LHC than at RHIC. However, because the  energy dependence of the parton energy loss and the
less steep initial jet spectra, the suppression factors will increase with $p_{T}$ more strongly at LHC than at RHIC.

Because of the increased number of jet production in heavy-ion collisions at LHC energies, there are increased
possibility of larger $p_{T}$ hadron production from recombination of parton showers from independent jets. This
production mechanism will be more important than the shower-thermal and thermal-thermal parton recombination
that have been considered more relevant in heavy-ion collisions at the RHIC
energy \cite{Hwa:2002tu,Greco:2003xt,Fries:2003vb}. Such contributions from jet-jet parton recombination will
likely increase the hadron yield at moderate $p_{T}$ and increase the values of the suppression factor $R_{AA}$
at the LHC energies.

\section*{Acknowledgements}

This work was supported by NSFC of China under Project Nos. 10825523, 10875052 and Key
Grant No. 11020101060, and by MOE and SAFEA of China under Project No.
$PITDU-B08033$, and by Key Laboratory of $Quark$ \& $Lepton$
Physics (Huazhong Normal University) of MOE of China under Project No.
QLPL200913, and by the Director, Office of Energy Research, Office of High Energy and Nuclear Physics, Divisions of Nuclear
Physics, of the U.S. Department of Energy under Contract No.
DE-AC02-05CH11231 and with the framework of the JET Collaboration,
and grant through Grant-in-Aid for Scientific Research No. 22740151 and
through the Excellent Young Researchers Oversea Visit Program (No. 213383)
of the Japan Society for the Promotion of Science. T.H. thanks members in the Nuclear Theory Program
at Lawrence Berkeley National Laboratory for kind hospitality during his sabbatical stay.

\end{document}